\documentstyle[twoside,fleqn,espcrc2,epsf]{article}

\newcommand{\lw}[1]{\smash{\lower2.ex\hbox{#1}}}
\def\simlt{\rlap{\lower 3.5 pt\hbox{$\mathchar \sim$}}\raise 1pt \hbox {$<$}}
\def\simgt{\rlap{\lower 3.5 pt\hbox{$\mathchar \sim$}}\raise 1pt \hbox {$>$}}

\title{Non-perturbative renormalization in domain-wall QCD
by Schr\"odinger functional scheme
\thanks{Poster presented by S.~Aoki}
}

\author{CP-PACS Collaboration:
  S.~Aoki\rlap,\address{Institute of Physics,
    University of Tsukuba, Tsukuba, Ibaraki 305-8571, Japan}
  Y.~Aoki\rlap,\address{RIKEN BNL Research Center,
    Brookhaven National Laboratory, Upton, NY 11973, USA}
  S.~Ejiri\rlap,\address{Department of Physics, University of Wales Swansea,
  Singleton Park, Swansea SA2 8PP, UK},
  M.~Fukugita\rlap,\address{Institute for Cosmic Ray Research,
    University of Tokyo, Kashiwa 277-8582, Japan}
  S.~Hashimoto\rlap,\address{High Energy Accelerator Research Organization
    (KEK), Tsukuba, Ibaraki 305-0801, Japan}
  N.~Ishizuka\rlap,$^{\rm a,}$\address{Center for Computational Physics,
    University of Tsukuba, Tsukuba, Ibaraki 305-8577, Japan}
  Y.~Iwasaki\rlap,$^{\rm a}$
  T.~Izubuchi\rlap,\address{Physics Department, Brookhaven National Laboratory,
  Upton, NY 11973, USA}
  K.~Kanaya\rlap,$^{\rm a}$
  T.~Kaneko\rlap,$^{\rm e}$
  Y.~Kuramashi\rlap,$^{\rm e}$
  M.~Okawa\rlap,$^{\rm e}$
  Y.~Taniguchi\rlap,$^{\rm a}$
  A.~Ukawa$^{\rm a,f}$ and
  T.~Yoshi\'e$^{\rm a,f}$
}
\begin{document}
\begin{abstract}
We formulate and numerically test the Schr\"odinger functional scheme 
for domain-wall QCD.
We then apply it to a non-perturbative calculation of the 
renormalization factors for vector and axial-vector currents 
in quenched domain-wall QCD with plaquette
and renormalization group improved gauge actions at $a^{-1}\simeq 2$ GeV.
\end{abstract}

\maketitle

\section{Introduction}
\label{sec:intro}
Recent calculations with domain-wall QCD (DWQCD) have shown that 
the good chiral property of domain-wall fermions
leads to good scaling behavior of physical observables
such as quark mass and $B_K$\cite{cppacs_bk}.
Aside from the quenched approximation,
the use of perturbative renormalization factors 
is the largest source of the uncontrolled systematic 
errors in these calculation.   Non-perturbative renormalization 
is required to reduce the total error to a few percent except 
for quenching, and we decide to employ the
Schr\"odinger functional(SF) scheme\cite{alpha},
which is a gauge-invariant and finite volume method, 
toward this goal.

In this report we formulate the SF scheme for DWQCD
and calculate the renormalization factors for 
vector and axial-vector currents
as a first step to obtain the renormalization factors 
for the quark mass and $B_K$.

\section{Schr\"odinger Functional in DWQCD}

We consider DWQCD\cite{FS} on an $L^3\times T\times N_s$
lattice.  The SF boundary condition for the domain-wall fermion 
is given by 
$\psi(\vec{x},x_0=0,s)= P_+ P_L(s)\rho(\vec{x})$,
$\bar\psi(\vec{x},x_0=0,s) =\bar\rho(\vec{x}) P_R(s)P_- $,
$\psi(\vec{x},x_0=T,s) = P_- P_L(s)\rho^\prime(\vec{x})$ and
$\bar\psi(\vec{x},x_0=T,s) =\bar\rho^\prime(\vec{x}) P_R(s)P_+$.
Here $\rho$, $\bar\rho$ are the boundary quark fields,
$P_L(s) = P_L\delta_{s,1}+P_R\delta_{s,N_s}$ and
$P_R(s) =P_R\delta_{s,1}+P_L\delta_{s,N_s}$
with $P_{R/L} =\frac{1}{2}(1\pm\gamma_5) $ and
$P_\pm =\frac{1}{2}(1\pm\gamma_0)$.

We follow the definitions and notations of ref.~\cite{sfZ}
for the calculation of the renormalization factors.  

\section{Tests of the formulation}

\begin{figure}[t]
\vspace{-0.5cm}
\centerline{\epsfxsize=6.8cm \epsfbox{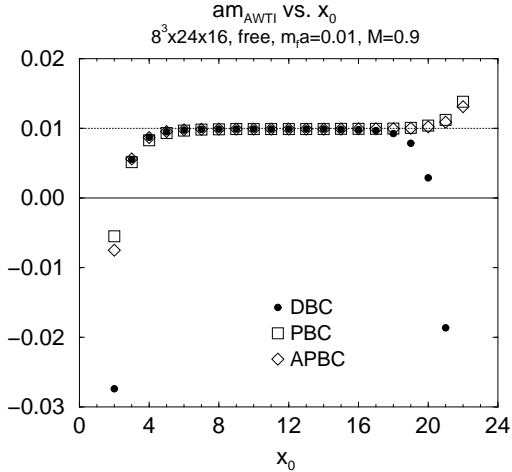}}
\vspace{-1.0cm}
\caption{$a m_{\rm AWTI}$ as a function of $x_0$ with 
Dirichlet(solid circles),
periodic(open squares) and anti-periodic(open diamonds) 
boundary conditions.}
\label{fig:mAWTI}
\vspace{-0.5cm}
\end{figure}

Since the SF boundary condition in time for $\rho=\bar\rho = 0$ is 
identical to the Shamir's domain-wall(Dirichlet) boundary 
condition in the fifth direction\cite{shamir},  
extra zero modes appear near $x_0 = 0$ and $T$.
We check whether these unwanted zero modes induce extra 
contribution to the low energy observables at $ 0 \ll x_0 \ll T$.
Here we consider the quark mass, $a m_{\rm AWTI}$, defined through 
the axial Ward-Takahashi identity(AWTI).
In Fig.~\ref{fig:mAWTI},
we plot $ a m_{\rm AWTI}$ for free theory as a function of $x_0$ with
Dirichlet, periodic and anti-periodic boundary conditions at the bare
quark mass $m_f a$=0.01, on an $8^3\times 24 \times 16$ lattice,
with the domain-wall height $M$=0.9.
The dependence of $a m_{\rm AWTI}$ on the boundary condition,
which is visible near the boundaries, disappears away from them.
Therefore we conclude that the extra zero modes associated with 
the Dirichlet boundary condition gives negligible effects 
to the determination of the renormalization 
factors evaluated at $x_0 \simeq T/2$.
Further analytic investigations on this problem for the free case will be 
given elsewhere\cite{aoki-kikukawa}.

\begin{figure}[t]
\vspace{-0.5cm}
\centerline{\epsfxsize=6.8cm \epsfbox{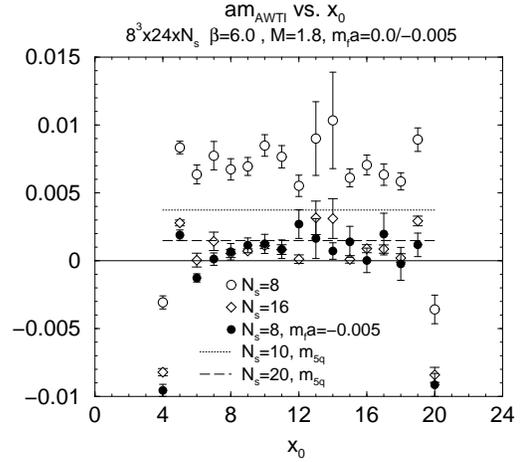}}
\vspace{-1.0cm}
\caption{$a m_{\rm AWTI}$ as a function of $x_0$ at $\beta\!=\!6.0$ on 
an $8^3\!\times\!24\!\times\!N_s$ lattice at 
$m_f a=0$(open symbols) and $-0.005$
(solid circles), together with $m_{5q}$\cite{chiral}.}
\label{fig:mq}
\vspace{-0.5cm}
\end{figure}

In Fig.~\ref{fig:mq}, $ a m_{\rm AWTI}$ for the SF scheme is plotted 
in quenched DWQCD at $\beta = 6.0$ of the plaquette action.
The $x_0$ dependence is also weak away from the boundaries.
Interestingly $m_{\rm AWTI}$ is non-zero even at $m_f a = 0$, 
and becomes smaller for larger $N_s$. Moreover
the value is consistent with $m_{5q}$,
the explicit breaking of chiral symmetry calculated from the conserved 
axial vector current of DWQCD\cite{chiral}.
Therefore 
$m_{\rm AWTI}$ for the SF scheme may be a better alternative 
as the measure of explicit chiral symmetry breaking in DWQCD, 
since it can be calculated directly at $m_f a = 0$ with much less
computational cost.
Note also that
the large explicit breaking in $m_{\rm AWTI}$ at $N_s=8$ 
(open circles) is well compensated if one takes a negative quark mass 
of $m_f=-0.005$ (filled circles).  Hence 
the domain-wall fermion  at $N_s\not=\infty$ 
may be considered as a highly improved Wilson fermion\cite{vpp}.

\begin{figure}[t]
\vspace{-0.5cm}
\centerline{\epsfxsize=6.8cm \epsfbox{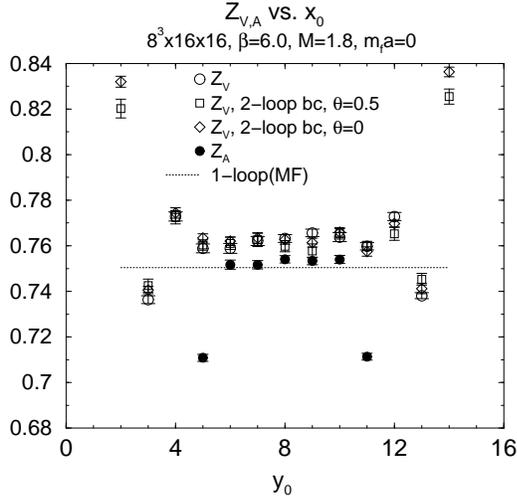}}
\vspace{-1.0cm}
\caption{$Z_{V,A}$ as a function of $x_0$ at $\beta=6$ on 
an  $8^3\times 16\times 16$ lattice with $M=1.8$ and $m_f a = 0$.
We compare the results from the boundary counter-terms 
at tree-level(circles) with those at 2-loop(squares and diamonds) as well
as those at $\theta = 0$ with that at $\theta = 0.5$(squares).}
\label{fig:zVA}
\vspace{-0.5cm}
\end{figure}

The non-perturbative renormalization factors for vector and axial-vector 
currents are defined by
$Z_V (1 + b_V m_f a) = f_1/ f_V (x_0) $ and
$Z_A^2 = 2f_1/f_{AA}(x_0, x_0^+, x_0^-)$,
where we fix $x_0^\pm = T/2\pm T/4$ and
put $ m_f a = 0$ for $Z_A$.
In Fig.~\ref{fig:zVA}, $Z_{V}$ and $Z_A$ are plotted as a function of $x_0$
at $\beta = 6.0$.
Similar to the case of $ a m_{\rm AWTI}$
a plateau is seen away from the boundaries. 
The relation $Z_V = Z_A$, exactly valid in perturbation theory,
is satisfied within 1--2\%.  Moreover the magnitude almost 
agrees with the mean-field(MF) improved one-loop value.
We also observe that $Z_V$ is insensitive to boundary parameters such as
the 2-loop boundary counter-terms for gauge fields and
the parameter $\theta$ of the twisted boundary condition for quarks.

\section{Results}

\begin{figure}[t]
\vspace{-0.5cm}
\centerline{\epsfxsize=6.6cm \epsfbox{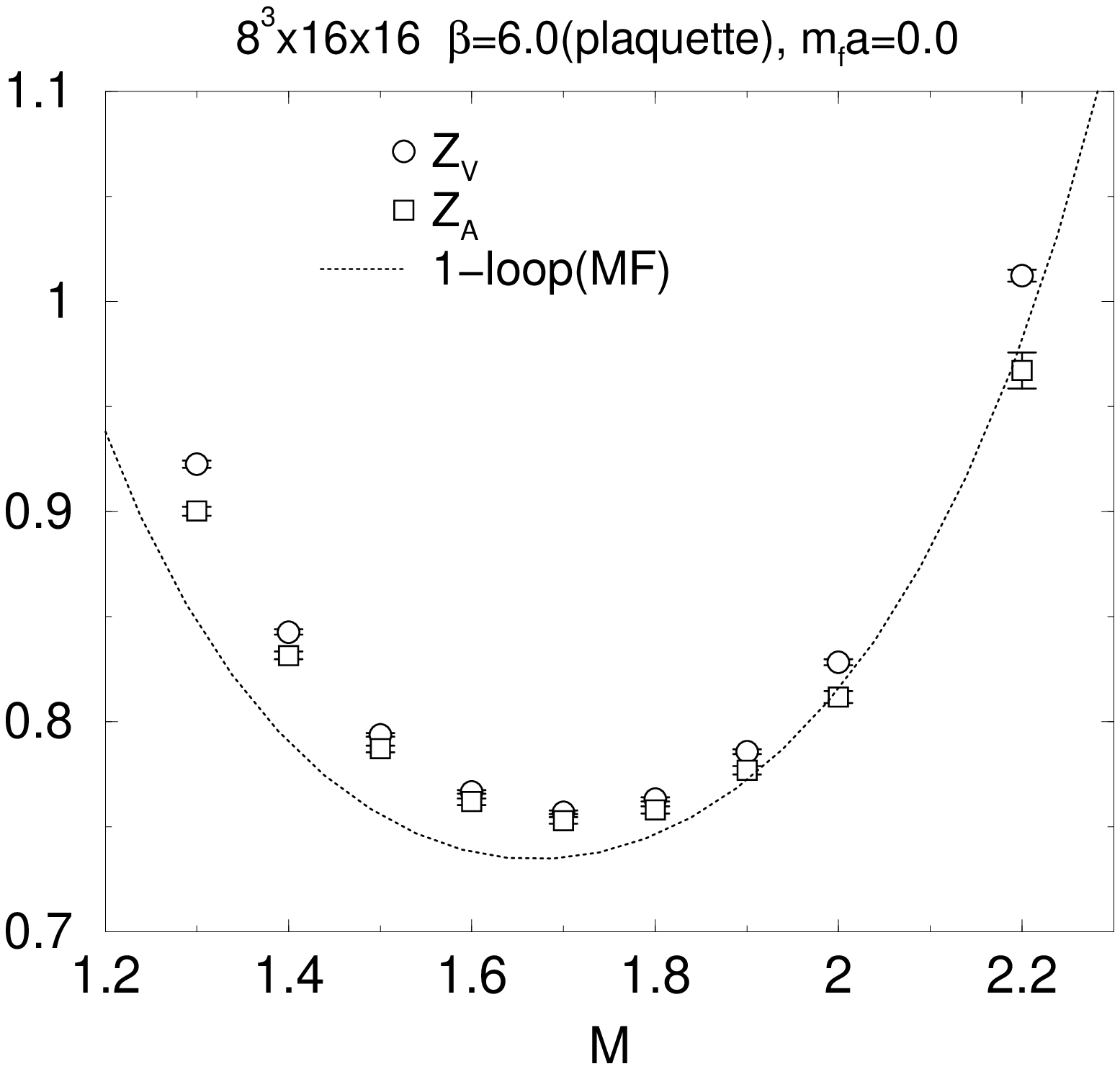}}
\centerline{\epsfxsize=6.6cm \epsfbox{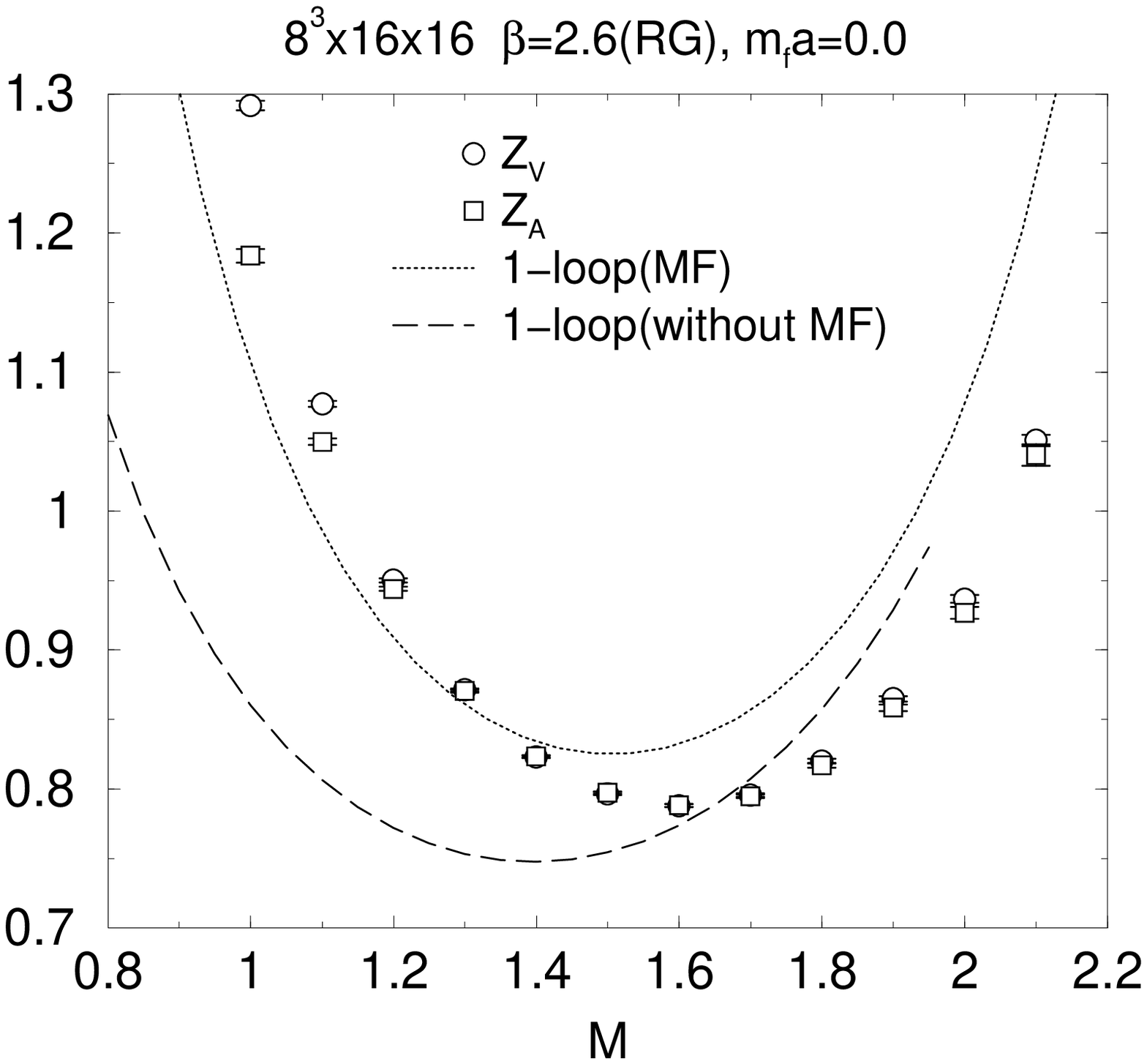}}
\vspace{-1.0cm}
\caption{$Z_{V}$ and $Z_A$ vs $M$ for plaquette(upper) and RG(lower) 
gauge actions.}
\label{fig:result}
\vspace{-0.7cm}
\end{figure}

We have calculated $Z_{V}$  and $Z_A$ in quenched DWQCD at 
$a^{-1} \simeq 2$ GeV with the plaquette action ($\beta = 6.0$) and 
with the renormalization group(RG) improved action ($\beta = 2.6$)
on an $8^3\times 16 \times 16$ lattice with $m_f a = 0$ for $M=1.0\sim 2.2$.
The results are summarized in Fig.~\ref{fig:result},
where $Z_V$ and $Z_A$ are plotted as a function of $M$, 
together with one-loop perturbative estimates with and without 
MF improvement.
For both gauge actions, $Z_V \simeq Z_A$ holds, and they show a minimum 
at $M\simeq 1.7$ for the plaquette action
or $M\simeq 1.6$ for the RG action.
Perturbative estimates without MF improvement 
fail, particularly for the plaquette action for which 
the curve can not be placed in the figure.
The agreement becomes much better with MF improvement for both actions.

\section{Future directions}

We are encouraged with the present results to proceed to an extension of the 
present work to scale-dependent cases such as quark masses
and four-quark operators needed for $B_K$. 

\vspace*{3mm}
S.A. thanks Profs. M.~L\"uscher, P.~Weisz and H.~Wittig
for useful discussion.
This work is supported in part by Grants-in-Aid of the Ministry of Education
(Nos.
11640294, 
12014202, 
12304011, 
12640253, 
12740133, 
13640260
13135204  
).

\end{document}